\documentclass[12pt]{article}   
\usepackage{amsmath, amsthm, amscd, amsfonts, amssymb, graphicx, color}
\usepackage[utf8]{inputenc}
\usepackage[toc,page]{appendix}
\usepackage{array}
\usepackage{multirow}
\usepackage{hyperref}
\usepackage{cite}
\usepackage[T1]{fontenc}
\usepackage{graphicx}
\usepackage{epstopdf} 
\usepackage{color}
\input{epsf}
\usepackage{lmodern}
\usepackage{wrapfig}
\usepackage{amsmath}
\newcommand{\be}{\begin{equation}}
\newcommand{\ee}{\end{equation}}
\newcommand{\bear}{\begin{eqnarray}}
\newcommand{\ear}{\end{eqnarray}}
\begin{document}
\title{Stationary states for a particle in a box with slanted walls\\ 
\vspace{.2cm}
\small Estados estacion\'arios de uma part\'{\i}cula numa caixa com paredes inclinadas}
\author{Nivaldo A. Lemos  \\
\small
{\it Instituto de F\'{\i}sica - Universidade Federal Fluminense}\\
\small
{\it Av. Litor\^anea s/n, Boa Viagem, 
24210-340, Niter\'oi - RJ, Brasil}\\
\small
{\it  nivaldolemos@id.uff.br}}

\date{\today}

\maketitle

\begin{abstract}

The quantum mechanical  problem of a particle in a infinite potential well with slanted walls is studied. The energy eigenvalues are determined by a transcendental equation involving the Airy function of the first kind. Two limiting cases are discussed. Next, the allowed energies  are found by   graphical and numerical methods. The graphical analysis hints that for large quantum numbers the consecutive energy levels get arbitrarily close together. This asymptotic  behavior of the energy spectrum actually  holds, as shown by means of an intuitive argument and confirmed by resorting to basic properties of Airy functions. In a sense, Bohr's correspondence  principle is more accurately fulfilled than in the cases of the standard particle in a box or the harmonic oscillator.  A state is regarded the more confined the less the probability of finding the particle in the classically forbidden region. For  some stationary states the confinement degree is calculated and a comparison is made with the harmonic oscillator, for which the ground state is the least confined of all stationary states.   In the symmetric case --- right and left walls of equal slope --- something  unexpected happens: beyond a certain critical slope the ground state is no longer the least confined eigenstate. Over and above its interesting physical features, this problem provides students with the opportunity to get acquainted with Airy functions, which do not belong to the standard mathematical repertoire of physics undergraduates.


\vspace{.2cm}

O  problema quanto-mec\^anico de uma part\'{\i}cula num po\c co de potencial infinito com paredes inclinadas \'e estudado. Os autovalores da energia s\~ao determinados por uma equa\c c\~ao transcendente envolvendo a fun\c c\~ao de  Airy de primeira esp\'ecie. Dois casos-limite s\~ao discutidos. Em seguida, as energias permitidas s\~ao encontradas por m\'etodos gr\'aficos e num\'ericos.  A an\'alise gr\'afica sugere que para grandes n\'umeros qu\^anticos os n\'{\i}veis de energia  consecutivos tornam-se arbitrariamente  pr\'oximos. Este comportamento assint\'otico  do espectro de energia  de fato se verifica, conforme mostrado por meio de um argumento intuitivo e confirmado pelo recurso  a   propriedades b\'asicas das fun\c c\~oes de Airy. Num certo sentido,  o princ\'{\i}pio da correspond\^encia de Bohr \'e realizado com mais precis\~ao do que nos  casos da part\'{\i}cula numa caixa padr\~ao e do oscilador harm\^onico.  Um  estado \'e considerado tanto mais confinado quanto menor a   probabilidade de encontrar a part\'{\i}cula   na regi\~ao  classicamente proibida. Para alguns estados estacion\'arios o grau de confinamento \'e  calculado e faz-se uma compara\c c\~ao com o osilador harm\^onico, para o qual o estado fundamental \'e o menos confinado de todos os estados estacion\'arios.   No caso sim\'etrico --- paredes esquerda e direita igualmente inclinadas --- algo inesperado acontece: acima de  uma certa inclina\c c\~ao cr\'{\i}tica o estado fundamental n\~ao \'e mais o autoestado menos confinado. Para 
al\'em de seus aspectos f\'{\i}sicos  interessantes, este problema propicia aos estudantes  a oportunidade de se familiarizar com as fun\c c\~oes de Airy, que n\~ao pertencem ao repert\'orio matem\'atico padr\~ao de graduandos em f\'{\i}sica.

\noindent{\bf Palavras-chave:} Estados ligados; Caixa com paredes inclinadas; Fun\c c\~oes de Airy; Mec\^anica Qu\^antica

\end{abstract}

\maketitle

\section{Introduction}\label{Intro}

 Confining potentials play a fundamental role in quantum mechanics, being essential for the  description  of particles trapped within a bounded region. Confined particles are characterized by a quantized  energy spectrum. The simplest of such potentials is the  infinite square well, also known as particle in a box, for which the time-independent Schr\"odinger equation is easily solved exactly. Although very simple, this model, either in  its one-dimensional or three-dimensional version,  finds significant applications to chemistry \cite{Autschbach}, quantum dots \cite{Dots}, semiconductor nanocrystals \cite{Kippeny}
and nuclear structure \cite{Krane}, to name just a few.

The next simplest confining potential is the harmonic oscillator parabolic potential energy. A hybrid of infinite well and parabolic potential is the so-called ``bathtub'' potential, which consists of an infinite well with parabolic walls \cite{Mazzitelli,Zou}. Suitable to modelling the confinement of electrons in nanostructures, this potential has been applied to the description of some important systems in condensed matter physics \cite{Crook,Martin}. A variant, which does not seem to have been explored before, 
 is the infinite well with a linearly rising potential on each side, or box with slanted walls. Here we undertake the study of the stationary states for a particle in such a potential. The  stationary wave functions are expressed in terms the Airy function of the first kind. The associated  energies are determined by a transcendental equation  which  is solved by graphical and numerical methods.  There is an infinite discrete set of bound states whose consecutive energies come indefinitely closer to each other for arbitrarily large quantum numbers. This behavior of the energy spectrum is quantitatively explained by way of an intuitive argument, whose conclusion is confirmed by taking advantage of some of the main properties of  Airy functions. Owing to the structure of the energy spectrum for large quantum numbers, it is fair to say that Bohr’s correspondence principle is more precisely satisfied than in the cases of the standard particle in a box or the harmonic oscillator.  A state is said to be the more confined the less the probability
that the particle penetrate into the classically forbidden region (tunneling probability). The degree of confinement is computed for a few stationary states and a comparison is made
with the harmonic oscillator, for which the ground state is the least confined of all energy eigenstates.  In the symmetric case — right
and left walls of equal slope — something a bit surprising happens: beyond a certain critical
slope the ground state ceases to be the least confined stationary state.

In Section \ref{Slanted}, the problem is stated and solved in terms of the Airy function of the first kind and its derivative. The transcendental equation whose solutions furnish the possible energies is derived and two significant limiting cases are addressed.
In Section \ref{graphical},  the  graphical as well as numerical determination of the energy eigenvalues is accomplished. An intuitive physical argument is invoked to explain the behavior of the energy spectrum for large quantum numbers, which is corroborated from the known properties of  Airy functions. In Section \ref{Normalization}, the stationary-state wave functions are normalized. The symmetric case of equally slanted walls is especially studied: the tunneling probability is computed for a few stationary states and a comparison is made with the harmonic oscillator. It is shown that there is a critical wall slope above which the ground state loses its position of least confined eigenstate. Section \ref{Conclusion} is dedicated to a few concluding remarks. The Appendix summarizes the features of Airy functions that are used in the course of the analysis.

\section{Slanted-wall one-dimensional box in quantum mechanics}\label{Slanted}

%\begin{figure}[h!]
%\begin{center}
% \includegraphics[width=.4\textwidth]{Semi-Infinite_EPS3}
%\caption{Semi-infinite potential energy square well. The particle is excluded from the region $x < 0$ by an impenetrable wall of infinitely high potential energy. }
%\label{Figure-potential}
%\end{center}
%\end{figure}

\begin{wrapfigure}{r}{0.35\textwidth}
\vspace{-30pt}
\begin{center}
\includegraphics[width=0.30\textwidth]{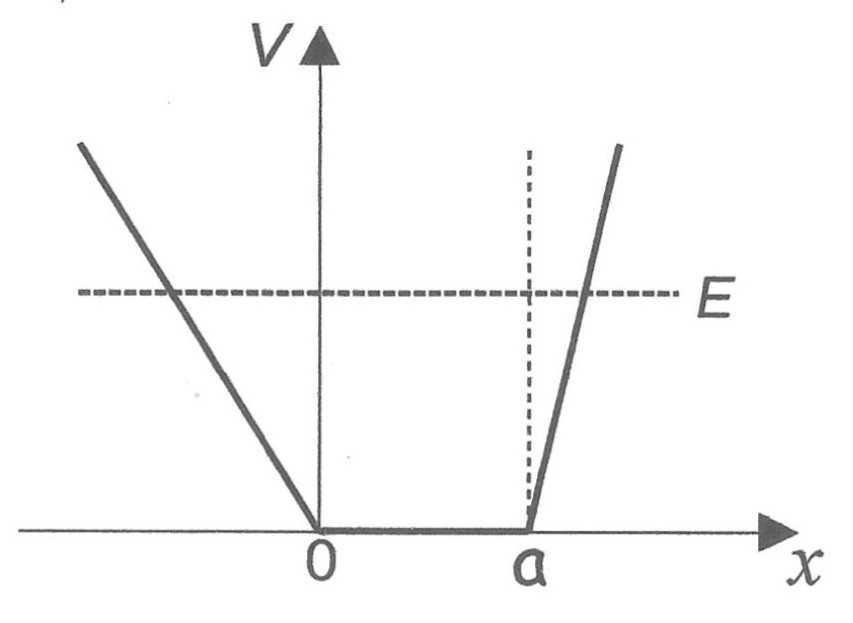}
\vspace{-10pt}
\end{center}
\vspace{-17.5pt}
\caption{\small Infinite potential well with slanted walls. }
\label{Figure-potential}
\end{wrapfigure}

Consider a particle of mass $m$ subject to the potential energy defined by
 \begin{equation}
\label{potential}
 V(x) = \left\{ \begin{array}{cl}
                    F_1 (x-a) &  \mbox{if \, $x>a$} \\
                       0  &  \mbox{if \, $0 \leq x \leq a$}\\
                     -F_2x & \mbox{if \, $x<0$}
                  \end{array} \right..
\end{equation}
 This potential energy is depicted in Fig. \ref{Figure-potential}. The width of the well's bottom is $a$; the positive constants $F_1, F_2$ have dimension of force. There can be definite energy bound states, described by square integrable wave functions, only if the energy $E$ is positive, which is assumed from now on.

The one-dimensional time-independent Schr\"odinger equation reads
\begin{equation}
\label{Schrodinger}
 \frac{d^2 \psi}{dx^2} + \frac{2m(E-V)}{\hbar^2} \psi = 0.
\end{equation}
The boundary conditions on the physically acceptable (square-integrable) solutions to equation \eqref{Schrodinger} are
\begin{equation}
\label{Boundary}
\lim_{x\to \pm \infty} \psi (x) =0; \qquad \psi, \psi^{\prime} \,\, \mbox{continuous at}\,\, x=0 \,\, \mbox{and} \,\, x=a.
\end{equation}

Because the functional form of $V(x)$ changes at $x=0$ and $x=a$, the time-independent Schr\"odinger equation must be set up separately  for each of the three regions $x<0$, $0 \leq x \leq a$ and $x> a$.

\subsection{Physical solution in each region}

$\blacksquare\,\,$ Region $x>a$. Since $V(x) = F_1(x-a)$, equation \eqref{Schrodinger} becomes
\begin{equation}
\label{Schrodinger-x-larger-a}
 \frac{d^2 \psi}{dx^2} + \frac{2mF_1}{\hbar^2}\Bigl(-x+a + \frac{E}{F_1} \Bigr) \psi = 0.
\end{equation}
It is useful to introduce the constant 
\begin{equation}
\label{k1}
 k_1 = \bigg( \frac{2mF_1}{\hbar^2}\bigg)^{1/3},
\end{equation}
which has dimension of (length)$^{-1}$, and the dimensionless variable
\begin{equation}
\label{dimensionless-xi}
 \xi  = k_1 \Bigl(x - a - \frac{E}{F_1} \Bigr).
\end{equation}
A simple use of the chain rule shows that in terms of $\xi$ equation \eqref{Schrodinger-x-larger-a} takes the cleaned-up form
\begin{equation}
\label{Schrodinger-x-larger-a-xi}
 \frac{d^2 \psi}{d\xi^2} - \xi \psi = 0.
\end{equation}
This is Airy's differential equation for $\psi(\xi)$ --- see Eq. \eqref{Airy-diff-equation-appendix} in the Appendix.
Two linearly independent solutions to this equation are the Airy functions $\mbox{Ai}(\xi)$ and $\mbox{Bi}(\xi)$, whose basic properties are described in the Appendix.
These functions are such that  $\mbox{Ai}(\xi) \to 0$ but  $\mbox{Bi}(\xi) \to \infty$ as $\xi \to \infty$ --- see equations \eqref{asymptptic-infinity-Ai-appendix} and \eqref{asymptptic-infinity-Bi-appendix}. Therefore, since $\xi \to \infty$ as $x \to \infty$, the physically acceptable solution to Eq. \eqref{Schrodinger-x-larger-a-xi} is $\psi (\xi) = C_1 \mbox{Ai}(\xi)$, where $C_1$ is an arbitrary constant. In terms of the original position variable we have 
\begin{equation}
\label{psi-x-larger-a}
 \psi (x) =  C_1 \mbox{Ai}\bigl( k_1(x-a-E/F_1)\bigr), \qquad  x>a.
\end{equation}

\medskip

$\blacksquare\,\,$ Region $0 \leq x \leq a$. Given that $V=0$, equation \eqref{Schrodinger} takes the form
\begin{equation}
\label{Schrodinger-inside}
 \frac{d^2 \psi}{dx^2} + k^2 \psi = 0
\end{equation}
where
\begin{equation}
\label{k}
k = \frac{\sqrt{2mE}}{\hbar}.
\end{equation}
With $C_2$ and $C_3$  arbitrary constants, the general solution to equation \eqref{Schrodinger-inside} is
\begin{equation}
\label{psi-inside-general}
 \psi (x) = C_2 \cos kx + C_3 \sin kx, \qquad 0 \leq x \leq a.
\end{equation}

\medskip

$\blacksquare\,\,$ Region $x < 0$. Inasmuch as  $V(x)=-F_2x$, equation \eqref{Schrodinger} takes the form
\begin{equation}
\label{Schrodinger-x-negative}
 \frac{d^2 \psi}{dx^2} + \frac{2mF_2}{\hbar^2}\Bigl(x+ \frac{E}{F_2} \Bigr) \psi = 0.
\end{equation}
Defining 
\begin{equation}
\label{k2-eta}
 k_2 = \Bigl( \frac{2mF_2}{\hbar^2}\Bigr)^{1/3},
\end{equation}
the introducion of the new variable
\begin{equation}
\label{eta}
 \eta  = - k_2 \Bigl( x + \frac{E}{F_2} \Bigr)
\end{equation}
reduces  equation \eqref{Schrodinger-x-negative} to
\begin{equation}
\label{Schrodinger-x-negative-eta}
 \frac{d^2 \psi}{d\eta^2} - \eta \psi = 0.
\end{equation}
Thus,
\begin{equation}
\label{psi-x-negative-esta-general}
 \psi (x) =  C_4 \mbox{Ai}(\eta) +  C_5 \mbox{Bi}(\eta).
\end{equation}
Since $\eta \to \infty$ as $x \to -\infty$, the presence of $ \mbox{Bi}(\eta)$ would lead to a non-normalizable wave function. Therefore one must set $C_5 =0$ to get the physically acceptable wave function
\begin{equation}
\label{psi-x-negative}
 \psi (x) =  C_4 \mbox{Ai}\bigl( -k_2(x + E/F_2)\bigr), \qquad  x<0.
\end{equation}

\subsection{Continuity  conditions}

The requirement that $\psi$ and $\psi^{\prime}$ be continuous at $x=0$ yields
\begin{eqnarray}
\label{continuity-x=zero}
 C_4 \mbox{Ai}(-k_2E/F_2) & = & C_2,\\
\label{continuity-derivative-x=zero}
-k_2C_4 \mbox{Ai}^{\prime}(-k_2E/F_2) & = & kC_3.
\end{eqnarray} 
Similarly, the condition that  $\psi$ and $\psi^{\prime}$ be continuous at $x=a$ furnishes 
\begin{eqnarray}
\label{continuity-x=a}
 C_2 \cos ka + C_3 \sin ka & = & C_1 \mbox{Ai}(-k_1E/F_1) ,\\
\label{continuity-derivative-x=a}
-kC_2 \sin ka + kC_3 \cos ka & = & k_1C_1 \mbox{Ai}^{\prime}(-k_1E/F_1).
\end{eqnarray}

Let us introduce the following positive dimensionless quantities:
\begin{equation}
\label{dimensionless-zs}
 z = ka, \qquad z_1 = k_1a, \qquad z_2 = k_2a.
\end{equation}
Together with  \eqref{k} this gives 
\begin{equation}
\label{Energy-in-terms-z}
 E = \frac{z^2\hbar^2}{2ma^2}.  
\end{equation} 
 With the use of \eqref{k1} and \eqref{k2-eta} in combination with \eqref{dimensionless-zs} one readily finds
\begin{equation}
\label{z1-tilde-in-terms-z}
 \frac{k_1E}{F_1} = \frac{z^2}{z_1^2}, \qquad   \frac{k_2E}{F_2} = \frac{z^2}{z_2^2}. 
\end{equation} 

\subsection{Transcendental equation for the allowed energies}

Taking \eqref{dimensionless-zs}  and \eqref{z1-tilde-in-terms-z} into account, equations \eqref{continuity-x=zero} and \eqref{continuity-derivative-x=zero} give rise to
\begin{equation}
\label{C2-over-C3}
 \frac{C_2}{C_3} = - \frac{z}{z_2} \frac{\mbox{Ai}(-z^2/z_2^2)}{\mbox{Ai}^{\prime}(-z^2/z_2^2)}. 
\end{equation} 
In their turn, equations \eqref{continuity-x=a} and \eqref{continuity-derivative-x=a} can be easily solved for $C_2$ and $C_3$ in the form
\begin{eqnarray}
\label{C2-solution}
 C_2 & = & z^{-1}\bigl[z\cos z \,\mbox{Ai}(-z^2/z_1^2) - z_1 \sin z \,\mbox{Ai}^{\prime}(-z^2/z_1^2) \bigr] C_1, \\
\label{C3-solution}
 C_3 & = & z^{-1}\bigl[ z\sin z \, \mbox{Ai}(-z^2/z_1^2) + z_1 \cos z \,\mbox{Ai}^{\prime}(-z^2/z_1^2) \bigr] C_1.
\end{eqnarray}
 As a consequence,
\begin{equation}
\label{C2-over-C3-again}
 \frac{C_2}{C_3} = \frac{z \cos z \,\mbox{Ai}(-z^2/z_1^2) - z_1 \sin z \,\mbox{Ai}^{\prime}(-z^2/z_1^2)}{z \sin z \, \mbox{Ai}(-z^2/z_1^2) + z_1 \cos z \,\mbox{Ai}^{\prime}(-z^2/z_1^2)}.
\end{equation}
In combination with equation \eqref{C2-over-C3} this yields
\begin{equation}
\label{transcendental}
 - \frac{z}{z_2} \frac{\mbox{Ai}(-z^2/z_2^2)}{\mbox{Ai}^{\prime}(-z^2/z_2^2)} = \frac{z \cos z \,\mbox{Ai}(-z^2/z_1^2) - z_1 \sin z \,\mbox{Ai}^{\prime}(-z^2/z_1^2)}{z \sin z \, \mbox{Ai}(-z^2/z_1^2) + z_1 \cos z \,\mbox{Ai}^{\prime}(-z^2/z_1^2)}.
\end{equation}
Since $z_1$ and $z_2$ are known, this transcendental equation determines the possible values of $z$. Then  the energy levels are found by means of equation \eqref{Energy-in-terms-z}.

\subsection{Limiting cases}\label{limiting-cases}

There are two immediate limiting cases of interest.

{\bf Case 1:}  $F_1,F_2 \to \infty$. In this limit the potential energy becomes the one for the particle in a standard  box, namely with vertical walls. From \eqref{k1}, \eqref{k2-eta} and \eqref{dimensionless-zs} it follows that $z_1,z_2 \to \infty$. The Airy function $\mbox{Ai}$ is infinitely differentiable and such that both $\mbox{Ai}(0)$ and $\mbox{Ai}^{\prime}(0)$ are nonzero (see Appendix). Therefore, in the limit $z_1,z_2 \to \infty$ equation \eqref{transcendental} narrows down to
\begin{equation}
\label{transcendental-limit-box}
 \tan z =0 \quad \Longrightarrow \quad \sin z = 0 \quad \Longrightarrow \quad  z = n\pi \,\,\,( n=1,2,3, \ldots) . 
\end{equation}
Then Eq. \eqref{Energy-in-terms-z} gives for the allowed energies
\begin{equation}
\label{energies-limit-box}
 E_n = \frac{n^2\pi^2\hbar^2}{2ma^2}. 
\end{equation}
These are the well-known energy eigenvalues for a particle in a one dimensional box. Given that  $k_1,k_2 \to \infty$ in such a way that $k_1/F_1, k_2/F_2 \to 0$ as $F_1,F_2 \to \infty$, and taking into account that  $\mbox{Ai}(x) \to 0$ as $x \to \infty$, equations \eqref{psi-x-larger-a} and \eqref{psi-x-negative} show that the limiting wave function vanishes outside the box, as expected. 

{\bf Case 2:}  $F_1\to 0$ or  $F_2 \to 0$. Then the particle is free either for $x<0$ or $x>a$. Let us first consider the case $F_2 \to 0$, which implies that $z_2 \to 0$ and $-z^2/z_2^2 \to -\infty$. Equations \eqref{asymptptic-minus-infinity-Ai-appendix} and  \eqref{asymptptic-minus-infinity-Aiprime-appendix} in the Appendix lead to
%\begin{eqnarray}
%\label{asymptptic-minus-infinity-Ai}
%\mbox{Ai}(x) &  \underset{x \to -\infty}{\xrightarrow{\hspace{1.2cm}}} & \frac{1}{\sqrt{\pi} \,\vert x \vert^{1/4}} \sin \Bigl( \frac{2}{3}\vert x\vert^{3/2} + \frac{\pi}{4}\Bigr),\\
%\label{asymptptic-minus-infinity-Aiprime}
%\mbox{Ai}^{\prime}(x) &   \underset{x \to -\infty}{\xrightarrow{\hspace{1.2cm}}}  & -\frac{\vert x \vert^{1/4}}{\sqrt{\pi} } \cos \Bigl( \frac{2}{3}\vert x\vert^{3/2} + \frac{\pi}{4}\Bigr).
%\end{eqnarray}
the following asymptotic behavior for the left-hand side of Eq. \eqref{transcendental}:
\begin{equation}
\label{transcendental-left-hand-side}
 - \frac{z}{z_2} \frac{\mbox{Ai}(-z^2/z_2^2)}{\mbox{Ai}^{\prime}(-z^2/z_2^2)} \underset{z_2 \to 0}{\xrightarrow{\hspace{1cm}}} 
-\cot \Bigl( \frac{2}{3} \, \frac{z^3}{z_2^3} - \frac{\pi}{4} \Bigr).
\end{equation}
Since $\cot u$ has no limit as $u\to \infty$, it follows that the limit of the  left-hand side of equation \eqref{transcendental} as $z_2 \to 0$ does not exist, implying that equation \eqref{transcendental} has no solution for $z$ in this limit. This is exactly as it was supposed to  be: there is no square-integrable state with definite energy if the particle is free for $x <0$. The same is true in the limit  $F_1 \to 0$ because the limit of the right-hand side of  equation \eqref{transcendental}  as $z_1 \to 0$ does not exist either, as can be checked with just a little bit more algebra.

\begin{figure}[t!]
\begin{center}
\includegraphics[width=.65\textwidth]{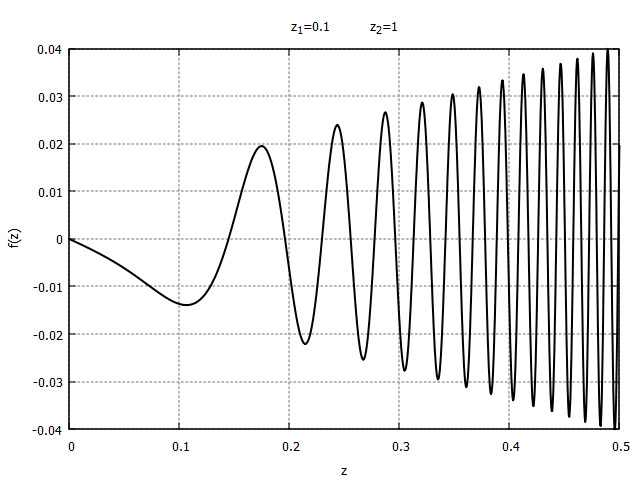}
\caption{\small Graph of the function $f(z)$ defined by equation \eqref{transcendental-no-denominator} for $z>0$ in the case $z_1=0.1$, $z_2=1$. The positive zeros of $f(z)$ determine the allowed energy levels. The consecutive zeros  of $f(z)$, of which there are infinitely many, get arbitrarily close to each other as $z  \to \infty$. Rounded to the tenth decimal place, the first five positive zeros of $f(z)$ are 
$0.1449448405, \,\, 0.1962267271,  \,\,  0.2298455404, \,\, 0.2559074516,  
 \,\, 0.2775989839$. }
\label{Figure-z1=01-z2=1}
\end{center}
\end{figure}

\section{Energy spectrum by graphical and numerical methods}\label{graphical} 

It is hopeless to find exact solutions to equation \eqref{transcendental}.  
Therefore, one must resort to  numerical, graphical or approximate algebraic methods for solving it.
It is clear that the physical solutions to Eq. \eqref{transcendental} are the positive zeros of the function $f: \mathbb{R} \to \mathbb{R}$ defined by
\begin{eqnarray}
\label{transcendental-no-denominator}
f(z) & = &  z\mbox{Ai}(-z^2/z_2^2) \bigl[ z \sin z \, \mbox{Ai}(-z^2/z_1^2) + z_1 \cos z \,\mbox{Ai}^{\prime}(-z^2/z_1^2)\bigr] \nonumber \\
&  &  + z_2 \mbox{Ai}^{\prime}(-z^2/z_2^2) \bigl[ z \cos z \,\mbox{Ai}(-z^2/z_1^2) - z_1 \sin z \,\mbox{Ai}^{\prime}(-z^2/z_1^2)\bigr].
\end{eqnarray}
For the purpose of finding graphical solutions for the energy levels, this function is much more convenient because it avoids the troublesome denominators that appear in \eqref{transcendental}, which vanish infinitely often. The energy eigenvalues are determined --- via equation \eqref{Energy-in-terms-z} --- from the  $z$ values for which the graph of $f(z)$ intersects the positive $z$-axis.\footnote{All graphical and numerical computations have been performed with the free software Gnuplot, in which the built-in $\mbox{airy}(x)$ function gives $\mbox{Ai}(x)$ for real argument.
The derivative of $\mbox{Ai}(x)$, which is not contained in the software as a built-in feature, has been implemented by $\mbox{airyprime}(x)=\bigl(\mbox{airy}(x+h)-\mbox{airy}(x-h) \bigr)/(2h)$ with $h=10^{-6}$. The Airy function of the second kind is the built-in $\mbox{Bi}(x)$ function in Gnuplot.}  Figure \ref{Figure-z1=01-z2=1} shows numerous solutions for $z$ in the case $z_1=0.1$, $z_2=1$. As expected, there are infinitely many solutions. As $z \to \infty$, which means $E \to \infty$, the consecutive energy levels become ever closer to each other, making the energy spectrum look almost continuous. 

It is worth pointing out that, on physical grounds, the allowed energies depend symmetrically on $F_1, F_2$, that is, $E=E(F_1,F_2)=E(F_2,F_1)$. Indeed, by rearranging its terms it can readily be seen that the function $f(z)$ defined by  \eqref{transcendental-no-denominator} depends symmetrically on the parameters $z_1, z_2$.

%\subsection*{Large quantum numbers and correspondence principle}\label{correspondence}

    Let us attempt an intuitive explanation of the high-energy behavior of the energy levels.  For large $E$, Fig. \ref{Figure-potential} shows that classically the particle ``feels'' as if it were trapped in a box of effective width
\begin{equation}
\label{effective-width}
 a_{\mbox{\scriptsize eff}} = a + \frac{E}{F_1} + \frac{E}{F_2} \approx \frac{E}{F}, \quad F=\frac{F_1F_2}{F_1+F_2}.
\end{equation}		
Then, equation \eqref{energies-limit-box}	leads one to estimate that	
\begin{equation}
\label{energies-limit-box-estimate}
 E_n \approx \frac{n^2\pi^2\hbar^2}{2ma_{\mbox{\scriptsize eff}}^2} \propto \frac{n^2}{E_n^2} \, \Longrightarrow \, E_n \propto n^{2/3}. 
\end{equation}		
For large $n$, with the help of the binomial expansion one finds that the difference between consecutive	energy levels  is
\begin{equation}
\label{difference-large-energies}
 \Delta E_n = E_{n+1} -E_n \propto (n+1)^{2/3}-n^{2/3} =n^{2/3}\bigg[ \Bigl( 1+ \frac{1}{n}\Bigr)^{2/3}-1\bigg] = \frac{2}{3n^{1/3}} + \mathcal{O}\bigl(\frac{1}{n^{4/3}}\bigr),
\end{equation}	
which tends to zero as $n \to \infty$.

Let us confront this with what the transcendental equation \eqref{transcendental} has to tell us for large $z$. This is essentially the same situation as the Case 2 considered in Subsection \ref{limiting-cases}. With the use of  \eqref{asymptptic-minus-infinity-Ai-appendix} and  \eqref{asymptptic-minus-infinity-Aiprime-appendix}, equation \eqref{transcendental} reduces to
\begin{equation}
\label{transcendental-right-hand-side} 
-\cot \Bigl( \frac{2}{3} \, \frac{z^3}{z_2^3} - \frac{\pi}{4} \Bigr) = \cot \Bigl( \frac{2}{3} \, \frac{z^3}{z_1^3} - \frac{\pi}{4} + z \Bigr) .
\end{equation}
Since the cotangent is an odd function with period $\pi$, this equation implies that
\begin{equation}
\label{values-large-z} 
 \Bigl(\frac{1}{z_1^3}  + \frac{1}{z_2^3}\Bigr) \, \frac{2z^3}{3} = n\pi, 
\end{equation}
 where the term linear in $z$ has been disregarded and it has also been taken into account that $n$ is large. This gives $z \propto n^{1/3}$ which, by \eqref{Energy-in-terms-z}, implies $E \propto n^{2/3}$, in confirmation of the intuitive semiclassical analysis.

Incidentally, the above intuitive argument works just fine for confining potentials such that $V(x) \propto \vert x \vert^{\alpha}$ for large $\vert x\vert$, where $\alpha$ is a positive constant. For high energies, the  ``box'' that traps the particle has the effective width $a_{\mbox{\scriptsize eff}} \propto E^{1/\alpha}$. The same reasoning predicated on equation \eqref{energies-limit-box-estimate} yields $E_n \propto n^{2\alpha/(\alpha + 2)}$. The particle in a box with slanted walls corresponds to  $\alpha =1$, implying that $E_n \propto n^{2/3}$, as already found. For the harmonic oscillator, $\alpha=2$ and it follows that  $E_n \propto n$, which entails equally spaced energy levels. The harmonic oscillator's uniqueness lies in that its large-quantum-number behavior extends throughout the  energy spectrum. If $0< \alpha <2$ then $0 < 2\alpha/(\alpha + 2) < 1$ and consecutive energy levels  come indefinitely close to each other as $n \to \infty$.

	\begin{figure}[t!]
\begin{center}
\includegraphics[width=.60\textwidth]{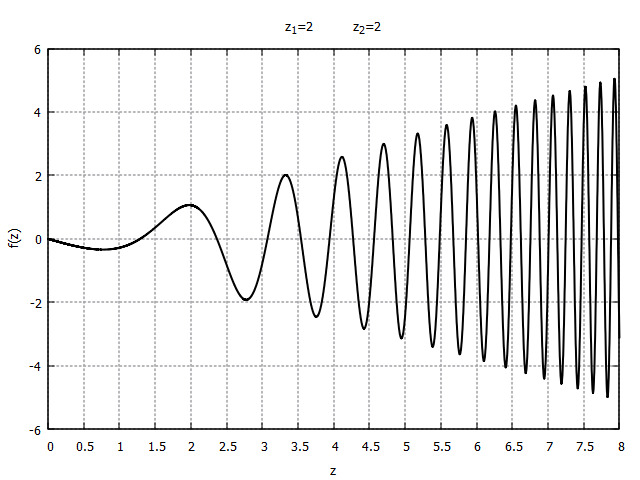}
\caption{\small The same as in  Fig. \eqref{Figure-z1=01-z2=1}, but  now for $z_1=z_2=2$.
%The permitted $z$ values are larger but, once again,  the difference between two consecutive solutions to $f(z)=0$ tends to zero as $z  \to \infty$. 
Rounded to the tenth decimal place, the first five positive zeros of $f(z)$ are 
$1.2898198725, \,\,  2.3644847297,  \,\,   3.0727839942, \,\,  3.5423302630, \,\,  3.9420924481  $. }
\label{Figure-z1=2-z2=2}
\end{center}
\end{figure}

There is something noteworthy about the potential energy \eqref{potential} as  far as the compliance with Bohr's  correspondence principle is concerned. For high energies, which is the same as high quantum numbers, the energy spectrum becomes effectively continuous, in conformity with classical mechanics.\footnote{Warning: the limit $\hbar \to 0$ is not necessarily the same as the limit of infinite quantum numbers \cite{Liboff}.} This is in contrast with the energy spectrum for both the particle in a box and the harmonic oscillator.  Of course the particle in a box and the harmonic oscillator obey Bohr's correspondence principle in the sense that the energy difference $\Delta E$ between consecutive energy levels is such that $\Delta E/E \to 0$ as $E \to \infty$. However, in the case of the particle in a box with slanted walls not only the energy difference between consecutive energy levels is small compared with the energy  but also $\Delta E$ itself tends to zero as $E \to \infty$, which makes the energy spectrum seem actually continuous and the particle act as more nearly free. It is useful to compare this behavior with that of the harmonic oscillator, whose energy spectrum is equally spaced. In the case of the harmonic oscillator the confining force on the particle becomes infinite at infinity,  whereas this force remains  constant for the particle in a box with slanted walls. As the energy grows bigger and bigger, the influence of a merely constant force on the particle becomes comparatively  less and less significant. Seen from this classical  perspective, the contrasting quantum behaviors that have just been described  look fairly plausible.

\begin{figure}[!ht]
    \centering
			\includegraphics[width=0.60\linewidth, height=0.3\textheight]{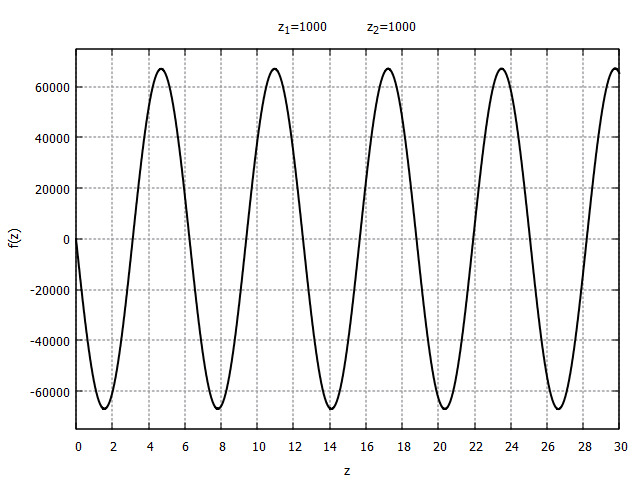}
		\caption{\small The same as in the two previous figures, but  now for  $z_1=z_2=1000$. The zeros  of $f(z)$ are  close to integer multiples of $\pi$, which is the exact result in the limit $z_1,z_2  \to \infty$.}
\label{Figure-z1=1000-z2=1000}		
\end{figure}

As $F_1, F_2$ increase, the walls get steeper and the potential well  effectively narrower. Since the particle becomes more localized, the  uncertainty principle  suggests that the allowed energies should become larger. Figure \ref{Figure-z1=2-z2=2} shows the solutions for $z$ in the case $z_1=z_2=2$. The $z$ values are indeed larger, but the overall behavior is the same as the one displayed in Fig. \ref{Figure-z1=01-z2=1}.  Finally, Fig. \ref{Figure-z1=1000-z2=1000} shows the solutions for $z$ in the case $z_1=z_2=1000$, which illustrates graphically the limit $z_1,z_2 \to \infty$. As shown in Subsection \ref{limiting-cases}, the positive zeros of $f(z)$ tend to $z_n=n\pi$, where $n$ is a positive integer.

In Table \ref{Eigenvalues} are listed the first five positive solutions to $f(z)=0$, in the symmetric case, for a few choices of $z_1=z_2$. These roots are needed for the computation of probabilities associated with the corresponding stationary states.  The exhibition of the first five zeros of $f(z)$ for the very special value $z_1=z_2= 2.55581$ will be justified below.

\begin{table}
\caption{\small First five $z$ values to ten decimal places in the symmetric case for six choices of $z_1=z_2$. These are solutions to $f(z)=0$ from which  the corresponding energy eigenvalues are found using equation \eqref{Energy-in-terms-z}. The  ground state is labeled with $n=0$. %The difference between consecutive roots of $f(z)$ diminishes as $n$ increases.
}
\begin{center}
\begin{tabular}{ |c||c|c|c|c|c|}
 \hline \hline
 $z_1=z_2$ & $n=0$ & $n=1$ & $n=2$ & $n=3$ & $n=4$ \\
 \hline
 1   & 0.7955111258  & 1.3575309823 &   1.6681556447 & 1.8953813141 & 2.0837723429 \\
  \hline
 2 &   1.2898198725  & 2.3644847297  &   3.0727839942  & 3.5423302630 & 3.9420924481 \\
\hline
 2.55581 &  1.4865367988 & 2.7905208546 &  3.7347178003 & 4.3596225896 & 4.8740812581  \\
\hline
 3  &  1.6165770040 & 3.0759769823  &  4.2014433303  & 4.9645505864 & 5.5691006885   \\
\hline
 4 &   1.8463998998  &  3.5820367860 &   5.0626332758  & 6.1659051005 & 6.9885397910   \\
\hline
 5 & 2.0162865038  & 3.9535453549  & 5.7057928708  & 7.1395134077 & 8.2231961014 \\
 \hline \hline
\end{tabular}
\end{center}
\label{Eigenvalues}
\end{table}

\section{Normalization and probabilities}\label{Normalization}

The normalization of the stationary-state wave functions requires
\begin{eqnarray}
\label{Normalization1} 
 \int_{-\infty}^{\infty} \vert \psi (x)\vert^2 dx & = & C_4^2 \int_{-\infty}^{0}\mbox{Ai}^2\bigl( -k_2(x + E/F_2)\bigr)\, dx + \int_0^a (C_2 \cos kx +C_3 \sin kx)^2 dx \nonumber \\
& & + C_1^2 \int_{a}^{\infty}\mbox{Ai}^2\bigl( k_1(x-a-E/F_1)\bigr)\, dx = 1,
\end{eqnarray}
where the constants can be chosen to be real. With the changes of variables $t=-k_2(x + E/F_2)$, $t=kx$ and   $t=k_1(x -a - E/F_1)$ for the first, second and third integrals, respectively, the previous condition becomes
\begin{eqnarray}
\label{Normalization2} 
  &  & \frac{C_4^2 a}{z_2}\int_{-z^2/z_2^2}^{\infty}\mbox{Ai}^2(t)\, dt +  \frac{C_1^2 a}{z_1}\int_{-z^2/z_1^2}^{\infty}\mbox{Ai}^2(t)\, dt  \nonumber \\
& &  + \frac{a}{4z} \bigl[4C_2C_3\sin^2 z +(C_2^2-C_3^2) \sin 2z + 2(C_2^2+C_3^2) z \bigr]  = 1,
\end{eqnarray}
where equations \eqref{dimensionless-zs} and \eqref{z1-tilde-in-terms-z} have been used. With the help of \eqref{integral-Ai-squared} this becomes
\begin{eqnarray}
\label{Normalization3} 
  &  & \frac{C_4^2 a}{z_2}\bigl[ (z^2/z_2^2)\mbox{Ai}^2(-z^2/z_2^2) + {\mbox{Ai}^{\prime}}^2(-z^2/z_2^2) \bigr] \nonumber \\
	& &  +  \frac{C_1^2 a}{z_1}\bigl[ (z^2/z_1^2)\mbox{Ai}^2(-z^2/z_1^2) + {\mbox{Ai}^{\prime}}^2(-z^2/z_1^2) \bigr] \nonumber \\
& &  + \frac{a}{4z} \bigl[4C_2C_3\sin^2 z +(C_2^2-C_3^2) \sin 2z + 2(C_2^2+C_3^2) z \bigr]  = 1.
\end{eqnarray}
By means of equations \eqref{continuity-x=zero}, \eqref{C2-solution} and \eqref{C3-solution} the above condition can be expressed in terms of $C_1$ alone, determining it and, consequently,  the three other constants.

As an illustration of the procedure, let us consider the ground state in the case $z_1=z_2=1$, for which  $z=0.7955111258$ as given in  Table \ref{Eigenvalues}. Equations 
\eqref{C2-solution}, \eqref{C3-solution} and \eqref{continuity-x=zero} yield
\begin{equation}
\label{C1-C2-C3-ground-state} 
  \frac{C_2}{C_1} =  0.5005101096, \quad  \frac{C_3}{C_1} = 0.2102893642, \quad  \frac{C_4}{C_1} = 1 .
\end{equation}
The third equality above could have been anticipated on symmetry grounds: the lowest energy wave function should be even with respect to $x=a/2$ because the same is true of the potential energy whenever $z_1=z_2$. With $n=0$ for the ground state, for odd-numbered stationary states the wave function is odd and one has $C_4=-C_1$, whereas for even-numbered stationary states the wave function is even and $C_4=C_1$. If $z_1 \neq z_2$ then $C_4 \neq \pm C_1$. Having made these remarks, which are useful for checking the numerical computations, it is time to move on with the procedure. Factoring out $C_1^2$ in \eqref{Normalization3}  and using \eqref{C1-C2-C3-ground-state} one
finds $C_1\sqrt{a}= 1.2377711820$. Thus we have
\begin{equation}
\label{C1-C2-C3-C4-ground-state} 
  C_1 = C_4 = \frac{1.2377711820}{\sqrt{a}}, \,\,\,   C_2 = \frac{0.6195169900}{\sqrt{a}}, \,\,\,  C_3 = \frac{0.2602901148}{\sqrt{a}}.
\end{equation}

\begin{figure}[!ht]
    \centering
			\includegraphics[width=0.65\linewidth, height=0.3\textheight]{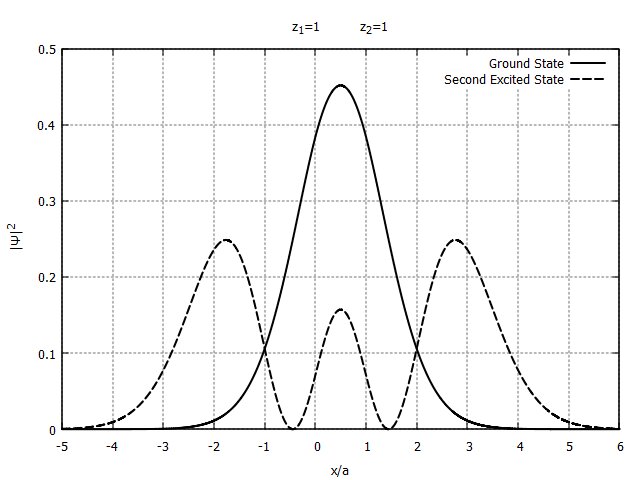}
		\caption{\small Position probability density  as a function of $x/a$ for the ground state  ($n=0$, solid line) and the second excited state ($n=2$, dashed line) in the case $z_1 = z_2 =1$.}
\label{Figure-z1=z2=1-Probability-density}		
\end{figure}

Figure \ref{Figure-z1=z2=1-Probability-density} shows a plot of the position probability density as a function of $x/a$  for the ground state and the second excited state in the case $z_1 = z_2 =1$. As expected, the probability distribution is symmetric about $x=a/2$. In terms 
of $x/a$,  the classically allowed region\footnote{In the general case,  in terms of $x/a$ the classically allowed region for the stationary state with ``energy'' $z$ is the interval $[-z^2/z_2^3, 1+z^2/z_1^3]$, as can be readily checked.}  for the ground state is the interval $[-0.63,1.63]$. For the second excited state the classically allowed region is  $[-2.78,3.78]$. It can be qualitatively seen that, as compared with the second excited state,  the ground state probability density penetrates appreciably farther into its classically forbidden region. This will be quantitatively substantiated presently.

\subsection{Tunneling probability}\label{class-allowed}
 
For a given energy $E$, the classically forbidden region is the union of the two disjoint regions  $x < -E/F_2$ and $x> a+E/F_1$. The tunneling probability, that is, the probability that the particle will be found in the classically forbidden region is 
\begin{equation}
\label{Ptunneling} 
 P_{\mbox{\scriptsize tunnel}}  =  C_4^2\int_{-\infty}^{-E/F_2}\mbox{Ai}^2(-k_2(x+E/F_2))\, dx + C_1^2\int_{a+E/F_1}^{\infty}
\mbox{Ai}^2(k_1(x-a-E/F_1))\, dx.
\end{equation}
Performing the changes of variables $t=-k_2(x+E/F_2)$ and $t=k_1(x-a-E/F_1)$ in the first and second integrals, respectively, and using equation \eqref{dimensionless-zs} one   arrives at 
\begin{equation}
\label{Ptunneling-explicit} 
 P_{\mbox{\scriptsize tunnel}}  =  \frac{C_4^2a}{z_2}\int_0^{\infty}\mbox{Ai}^2(t)\, dt + \frac{C_1^2a}{z_1}\int_0^{\infty}\mbox{Ai}^2(t)\, dt.
\end{equation}
Making use of equation \eqref{integral-Ai-squared} one finally gets 
\begin{equation}
\label{Ptunneling-final} 
P_{\mbox{\scriptsize tunnel}} = \Bigl(  \frac{C_1^2a}{z_1} +  \frac{C_4^2a}{z_2}\Bigr)\, {\mbox{Ai}^{\prime}}^2(0).
\end{equation}
It is worth noting that this probability implicitly depends on $z$, hence on the energy, through the constants $C_1$ and $C_4$.

The rows in Table \ref{Probabilities} give the tunneling probability in percentage for the first five stationary states in the symmetric case for some choices of $z_1=z_2$. The corresponding probabilities for the harmonic oscillator \cite{Diamond} are also presented in the last row for the sake of comparison.\footnote{It is worth pointing out that for large $n$ the formula $P(n) \approx 0.13397/(n+1/2)^{1/3}$ gives an excellent approximation to the probability that the particle will be found in the classically forbidden region for $n$-th eigenstate of the harmonic oscillator \cite{Diamond,Jadczyk}. Already for $n=4$ this asymptotic formula gives a result that differs from the exact value by less that $3\%$.} 
One can reasonably characterize  the degree of confinement of a quantum state by its tunneling probability as follows: the confinement degree is bigger the smaller the tunneling probability. Accordingly, as
far as the ground state is concerned, up to a little above $z_1=z_2=2$ the slanted-wall potential is less confining than the oscillator's parabolic potential.  This seems physically sensible: for small $z_1=z_2$ the walls are much too close to horizontal to significantly confine the particle. Nonetheless, for large $z_1=z_2$ the ground state becomes more confined than the oscillator's because
the walls become nearly vertical and the potential almost perfectly confines the particle to the bottom of the well, which for all intents and purposes is the classically allowed region. No surprise here.

\begin{table}[htb]
\caption{\small Tunneling probability to four decimal places (in percentage) for the first five stationary states in the symmetric case $z_1=z_2$. At approximately $z_1=z_2=2.55581$ there is a transition: the tunneling probabilities for the ground state and the first excited state become equal.  For bigger values of  $z_1=z_2$ the   tunneling probability ceases to be  maximal for the ground state. For the sake of reference, the last row displays the corresponding probabilities for the harmonic oscillator.}
\begin{center}
\begin{tabular}{ |c||c|c|c|c|c|}
 \hline \hline
 $z_1=z_2$ & $n=0$ & $n=1$ & $n=2$ & $n=3$ & $n=4$ \\
 \hline
 1   & 20.5260   & 13.6216 &   11.6070 & 10.3843 & 9.5523 \\
  \hline
 2 &   15.8453  & 13.5345  &  11.1985 & 10.3290 & 9.4221 \\
\hline
 2.55581 &  13.3437 & 13.3437 &  10.8901 & 10.2168 & 9.3535  \\
\hline
 3  &  11.5706 & 13.0771  &  10.6338  & 10.0656 & 9.3102   \\
\hline
 4 &   8.3807  &  12.1045 &  10.1090  & 9.5347 & 9.1807   \\
\hline
 5 & 6.1374  & 10.7791  & 9.6573  & 8.8759 & 8.8591 \\
\hline
Harm Osc & 15.7299 & 11.1610 & 9.5069 & 8.5482 & 7.8926 \\
 \hline \hline
\end{tabular}
\end{center}
\label{Probabilities}
\end{table}

\begin{figure}[!ht]
    \centering
			\includegraphics[width=0.65\linewidth, height=0.3\textheight]{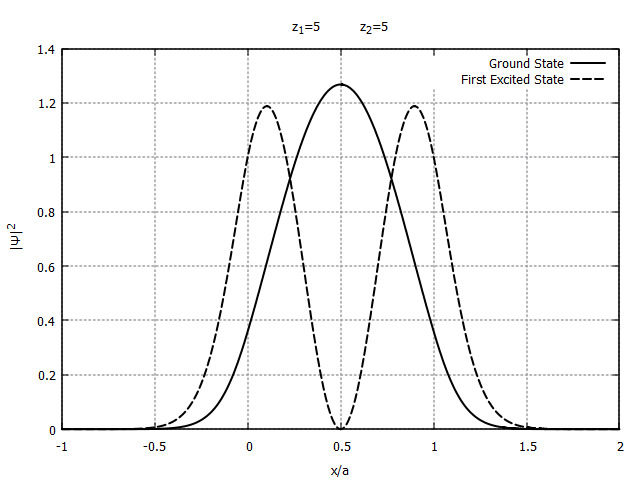}
		\caption{\small Position probability density  as a function of $x/a$ for the ground state  ($n=0$, solid line) and the first excited state ($n=1$, dashed line) in the case $z_1 = z_2 =5$.}
\label{Figure-z1=z2=5-Probability-density}		
\end{figure}

But something unexpected comes into view when one considers  the confinement degree of the ground state vis-\`a-vis the excited states. According to our confinement criterion, for a harmonic oscillator the ground state is the least confined of all stationary states, and the confinement degree steadily increases as the quantum number $n$ increases \cite{Diamond}.  This is not what happens in the case of the particle in a box with slanted walls. As shown in Table \ref{Probabilities}, at approximately $z_1=z_2=2.55581$ a transition takes place: the tunneling probabilities for the ground state and the first excited state become equal. For smaller values of $z_1=z_2$ the behavior is the same as that of the harmonic oscillator, namely the ground state is the least confined of all stationary states. But for values of $z_1=z_2$ larger than the transition value, it is the first excited state that becomes the least confined of all. Fig. \ref{Figure-z1=z2=5-Probability-density} illustrates the case $z_1=z_2=5$. In terms of $x/a$, for the ground state the classically allowed region is $[-0.03,1.03]$, whereas it is 
$[-0.13,1.13]$ for the first excited state. It is visually apparent that the first excited  state spreads considerably deeper into its classically forbidden region than it is case of the ground state. As Table \ref{Probabilities} shows, for $z_1=z_2=4$ and higher the first excited state remains the least confined of all, but a growing number of excited states get less confined than the ground state. I have not been able to find a physically intuitive explanation for this peculiar behavior.

\section{Conclusion}\label{Conclusion}

The infinite potential well with slanted walls is a confining potential that gives rise to some interesting behavior in quantum mechanics. The allowed energies, which are determined by a transcendental equation involving the Airy function of the first kind, have been found by graphical and numerical means.  It turns out that for high quantum numbers the energy spectrum becomes, for all practical purposes, continuous. It follows that Bohr's correspondence principle is more strongly fulfilled than in the cases of the standard particle in a box or the harmonic oscillator.   A state is deemed the more confined the less the tunneling probability or probability of leakage into the classically forbidden region. The symmetric potential well, whose walls are of equal slope, has been studied in more detail. For  a few stationary states the tunneling probability has been  calculated and a comparison  made with the harmonic oscillator, for which the ground state is the least confined of all eigenstates.   Something  unexpected has been found to happen: beyond a certain critical slope the ground state is no longer the least confined eigenstate. 

In addition to being attractive in virtue of its physical features, this problem allows undergraduate physics students to familiarize themselves with Airy functions, thus enriching their mathematical toolkit.

\appendix
\renewcommand{\theequation}{A.\arabic{equation}}
\setcounter{equation}{0}
\section*{Appendix: Basic properties of Airy functions}
 All results quoted in this Appendix have been collected from \cite{Airy}, a fine reference on Airy functions and their applications to classical and quantum physics.

Airy's differential equation for $y(x)$ is
\begin{equation}
\label{Airy-diff-equation-appendix}
 y^{\prime\prime} - x y = 0.
\end{equation}
Two linearly independent solutions to this equation are the Airy functions of the first and second kind, respectively  $\mbox{Ai}(x)$ and $\mbox{Bi}(x)$. These functions admit the following  integral representations: 
 
\begin{eqnarray}
\label{Airy-Ai-integral-rep}
 \mbox{Ai}(x) & = & \frac{1}{\pi} \int_0^{\infty} \cos \Bigl(\frac{t^3}{3} +tx \Bigr)\, dt, \\
\label{Airy-Bi-integral-rep}
 \mbox{Bi}(x) & = & \frac{1}{\pi} \int_0^{\infty} \bigg[ \exp\Bigl(  -\frac{t^3}{3} + tx\Bigr) + \sin \Bigl(\frac{t^3}{3} +tx \Bigr) \bigg]\, dt.
\end{eqnarray}

\begin{figure}[!ht]
    \centering
			\includegraphics[width=0.5\linewidth, height=0.3\textheight]{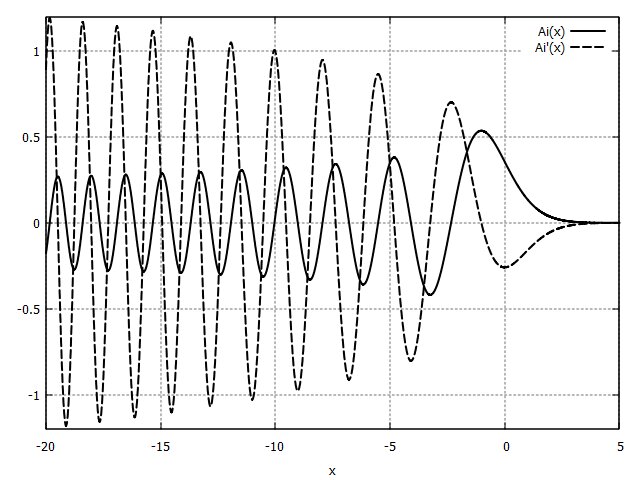}
        %\caption{$dt =$}
        %\label{fig:prob1_6_1}
		\caption{\small The Airy function of the first kind $\mbox{Ai}$ (solid line) and its derivative (dashed line).}
\label{Figure-Airy-Ai-and-derivative}		
\end{figure}

The Airy function of the first kind $\mbox{Ai}(x)$ and its derivative tend exponentially to zero as $x \to \infty$:
\begin{eqnarray}
\label{asymptptic-infinity-Ai-appendix}
  \mbox{Ai}(x)  & \underset{x \to \infty}{\xrightarrow{\hspace{0.8cm}}} & \frac{e^{-\frac{2}{3}x^{3/2}}}{2\sqrt{\pi} \, x^{1/4}},\\
\label{asymptptic-infinity-Aiprime-appendix}
\mbox{Ai}^{\prime}(x) &   \underset{x \to \infty}{\xrightarrow{\hspace{0.8cm}}}  & -\frac{x^{1/4}\, e^{-\frac{2}{3}x^{3/2}}}{2\sqrt{\pi}}.
\end{eqnarray}
As $x \to -\infty$, the Airy function $\mbox{Ai}(x)$ performs infinitely many oscillations with amplitude that slowly tends  to zero, whereas its derivative also oscillates infinitely often but with amplitude that slowly grows without bound:
\begin{eqnarray}
\label{asymptptic-minus-infinity-Ai-appendix}
 \mbox{Ai}(x) &  \underset{x \to -\infty}{\xrightarrow{\hspace{1.2cm}}} & \frac{1}{\sqrt{\pi} \,\vert x \vert^{1/4}} \cos \Bigl( \frac{2}{3}\vert x\vert^{3/2} - \frac{\pi}{4}\Bigr),\\
\label{asymptptic-minus-infinity-Aiprime-appendix}
\mbox{Ai}^{\prime}(x) &   \underset{x \to -\infty}{\xrightarrow{\hspace{1.2cm}}}  & \frac{\vert x \vert^{1/4}}{\sqrt{\pi} } \sin \Bigl( \frac{2}{3}\vert x\vert^{3/2} - \frac{\pi}{4}\Bigr).
\end{eqnarray}
The behavior of $\mbox{Ai}$ and its derivative is graphically shown in Fig. \ref{Figure-Airy-Ai-and-derivative}.

\begin{figure}[!ht]
\begin{center}
\includegraphics[width=.5\textwidth]{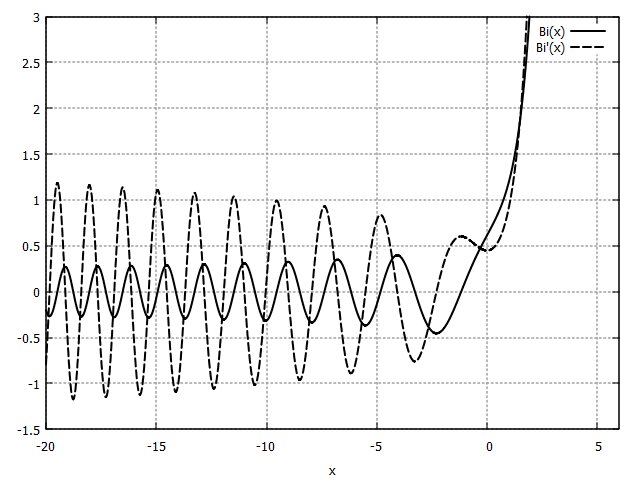}
\caption{\small The Airy function of the second kind $\mbox{Bi}$ (solid line) and its derivative (dashed line). }
\label{Figure-Airy-Bi-and-derivative}	
\end{center}
\end{figure}

The Airy function of the second kind $\mbox{Bi}(x)$ and its derivative grow exponentially as $x \to \infty$:
\begin{eqnarray}
\label{asymptptic-infinity-Bi-appendix}
  \mbox{Bi}(x)  & \underset{x \to \infty}{\xrightarrow{\hspace{0.8cm}}} & \frac{e^{\frac{2}{3}x^{3/2}}}{\sqrt{\pi} \, x^{1/4}},\\
\label{asymptptic-infinity-Biprime-appendix}
\mbox{Bi}^{\prime}(x) &   \underset{x \to \infty}{\xrightarrow{\hspace{0.8cm}}}  & \frac{x^{1/4}\, e^{\frac{2}{3}x^{3/2}}}{\sqrt{\pi}}.
\end{eqnarray}
Just like its first-kind counterpart, the Airy function $\mbox{Bi}(x)$ and its derivative oscillate infinitely often as $x \to -\infty$:
\begin{eqnarray}
\label{asymptptic-minus-infinity-Bi-appendix}
 \mbox{Bi}(x) &  \underset{x \to -\infty}{\xrightarrow{\hspace{1.2cm}}} & - \frac{1}{\sqrt{\pi} \,\vert x \vert^{1/4}} \sin \Bigl( \frac{2}{3}\vert x\vert^{3/2} - \frac{\pi}{4}\Bigr),\\
\label{asymptptic-minus-infinity-Biprime-appendix}
\mbox{Bi}^{\prime}(x) &   \underset{x \to -\infty}{\xrightarrow{\hspace{1.2cm}}}  & \frac{\vert x \vert^{1/4}}{\sqrt{\pi} } \cos \Bigl( \frac{2}{3}\vert x\vert^{3/2} - \frac{\pi}{4}\Bigr).
\end{eqnarray}
The general behavior of $\mbox{Bi}$ and its derivative is graphically shown in Fig. \ref{Figure-Airy-Bi-and-derivative}.

It should be noted that in a bounded domain both $\mbox{Ai}$ and $\mbox{Bi}$ are physically acceptable solutions to Airy's differential equation.
 
As shown by the asymptotic forms \eqref{asymptptic-minus-infinity-Ai-appendix}, \eqref{asymptptic-minus-infinity-Aiprime-appendix}, \eqref{asymptptic-minus-infinity-Bi-appendix}, \eqref{asymptptic-minus-infinity-Biprime-appendix}, and indicated by Figs. \ref{Figure-Airy-Ai-and-derivative} and \ref{Figure-Airy-Bi-and-derivative},  $\mbox{Ai}$ and $\mbox{Bi}$ together with their derivative have infinitely many negative zeros.

The Airy functions of the first and second kind as well as  their derivative do not vanish at the origin. Their exact and  numerical values (to ten decimal places) are
\begin{eqnarray}
\label{value-origin-Ai-Bi-appendix}
 & &\mbox{Ai}(0) = \frac{\mbox{Bi}(0)}{\sqrt{3}} = \frac{1}{3^{2/3} \Gamma\bigl( \frac{2}{3}\bigr)} = 0.3550280539,\\
\label{value-origin-Aiprime-Biprime-appendix}
& & \mbox{Ai}^{\prime}(0) = -\frac{\mbox{Bi}^{\prime} (0)}{\sqrt{3}} = -\frac{1}{3^{1/3} \Gamma\Bigl( \frac{1}{3}\Bigr)} = - 0.2588194038,
\end{eqnarray}
where $\Gamma$ is Euler's Gamma function.

Out of many known integrals involving Airy functions, the ones relevant for the purposes of the present study are the primitive
\begin{equation}
\label{integral-indefinite-Ai-squared}
 \int \mbox{Ai}^2(x) \, dx = x\mbox{Ai}^2(x) - {\mbox{Ai}^{\prime}}^2(x)
\end{equation}
and the consequent definite integral
\begin{equation}
\label{integral-Ai-squared}
 \int_x^{\infty} \mbox{Ai}^2(t) \, dt = -x\mbox{Ai}^2(x) + {\mbox{Ai}^{\prime}}^2(x).
\end{equation}

\end{document}